\documentclass[11pt,a4paper]{article}

\usepackage{amsfonts}
\usepackage[dvips]{graphicx}
\usepackage{amsthm}
\usepackage{epsfig}
\usepackage{t1enc}
\usepackage{amssymb}
\usepackage{latexsym}
\usepackage{bm}

\title{\bf A criterion for bubble formation
\\
in de Sitter universe}
\author{V. Balek\footnote{e-mail address: balek@fmph.uniba.sk}\ \ and
M.Demetrian\footnote{e-mail address: demetrian@sophia.dtp.fmph.uniba.sk}
\\
{\it Department of Theoretical Physics, 
Comenius University, Mlynsk\'a dolina}
\\
{\it 842 48 Bratislava, Slovakia}}
 
\begin{document}
\maketitle
\maketitle\abstract

{The influence of the shape of scalar field potential on the outcome
of vacuum decay in de~Sitter universe is studied. Sufficient condition 
for vacuum decay via bubble formation, described by Coleman - de Luccia 
instanton, is revisited and necessary condition
is found. Both conditions require that the curvature of the potential
is greater than 4 $\times$ (Hubble constant)$^2$, but while 
the sufficient condition states that this inequality is to be valid at the 
top of the barrier, the necessary condition requires that it holds 
at least somewhere throughout the barrier. The conditions 
leave a 'grey zone' in parameter space, which however seems
to be forbidden for quartic potential as well as for quadratic potential
with a narrow peak proposed by Linde in the context of open inflation.}



\vskip 5mm
\section{Introduction}
\vskip 1mm
The concept of vacuum decay via formation of rapidly expanding bubbles 
was introduced by Coleman \cite{col}, and its general relativistic extension, 
with bubbles created in de~Sitter universe, was developed by Coleman and 
de~Luccia \cite{cdl}. The process was an important ingredient of old inflation
\cite{old}, and it emerged again in the scenario of open inflation
\cite{{bgt},{lme},{toy}}.
An alternative mode of vacuum decay, applying presumably to the
initial stage of new inflation \cite{{new1},{new2}}, was proposed
by Hawking and Moss \cite{hmo}. In this mode, the transition 
should occur in the whole universe at once; however, as seen
from elementary considerations and confirmed by stochastic theory \cite{lst},
this must not be taken too literally. In fact, homogeneous transition takes 
place only inside the event horizon where the metric assumes de~Sitter form 
due to cosmological no-hair theorem.
\vskip 1mm
Each mode of vacuum decay is described by corresponding
instanton: the decay in a bubble by Coleman - de~Luccia instanton
(CdL instanton), and the decay in a horizon-size domain by Hawking -
Moss instanton (HM instanton). CdL instanton is
a squeezed 4-sphere with an O(4) configuration of scalar field living
on it which contains a 4D version of the bubble; HM instanton
is a round 4-sphere filled with scalar field which is
located at the top of the barrier everywhere.
It has been argued \cite{{hmo},{lys}} that CdL instanton exists only
if the curvature of the potential in the false vacuum is greater
than 4 $\times$ (Hubble constant)$^2$. The radius of the 4-sphere equals
approximately (Hubble constant)$^{-1}$ and the width of the bubble
wall equals approximately (curvature of the potential)$^{-1/2}$, thus
the criterion just states that the bubble must fit into the sphere.
If it does not, vacuum decays via HM instanton
which exists regardless of the shape of the potential.
\vskip 1mm
The criterion for the existence of CdL instanton has been formulated
as approximate, valid only for potentials of the kind used in new
inflation. However, it might be expected that this criterion, or some
modification of it, holds for an arbitrary potential as well.
In this article we propose a candidate for a generally valid criterion
and demonstrate its effectiveness on two examples.



\vskip 1mm
\section{Sufficient condition}
\label{sec:SC}
\begin{figure}[h]
\centerline{\includegraphics[width=8cm]{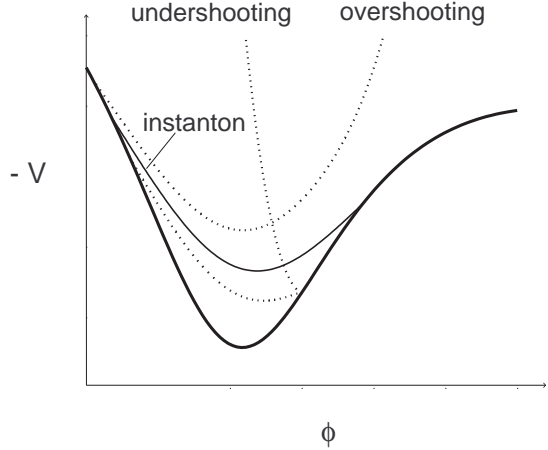}}
\caption{Energy curves}
\label{fig:En}
\end{figure}

\vskip 1mm
CdL instanton is a nontrivial finite-action O(4) solution 
of Eucleidean equations for coupled gravitational and scalar field, 
obtained in a
theory with an effective potential containing true vacuum (global
minimum in which it equals zero) and false vacuum (local minimum)
separated from the true vacuum by a finite potential barrier. It is given
by two functions $\phi (\rho)$ and $a (\rho)$, where $\phi$ is
the scalar field, $a$ is the radius of 3-spheres of homogenity determined
from the circumference and $\rho$ is the radius of 3-spheres of
homogenity measured from the centre. Denote the effective potential
$V$. Functions $\phi$ and $a$ obey the equations
\begin{equation}
\ddot \phi + \frac {3\dot a}a\ \dot \phi = V', \ \ \dot a^2 = \frac {8\pi}3
\left( \frac 12 \dot \phi^2 - V \right)\ a^2 + 1,
\label{eq:Eq12}
\end{equation}
where the dot denotes differentiation with respect to $\rho$ and the prime
denotes differentiation with respect to $\phi$. (Throughout the paper, we
use 'Mr. Tompkins' system of units in which $c = G = \hbar = 1$.)
Let us write down also the second order equation for $a$ which is
often helpful in general considerations, and is more appropriate
for numerical calculations than the first order equation. It reads
\begin{equation}
\ddot a = - \frac {8\pi}3 \left( \dot \phi^2 + V \right)\ a.
\label{eq:Eq2a}
\end{equation}
The function $a$ is concave everywhere and starts from zero at $\rho
= 0$ (the second possibility, namely that it is infinite there, is excluded 
by concavity), thus it must first increase and then decrease until it
reaches zero at some $\rho_f$. To ensure the finitness of action we
require that the solution is finite both at $\rho = 0$ and $\rho = \rho_f$.
The corresponding boundary conditions are
\begin{equation}
a(0) = \dot \phi (0) = \dot \phi (\rho_f) = 0.
\label{eq:BC}
\end{equation}
Denote $\phi_i$ and $\phi_f$ the initial and final value of $\phi$.
If we choose $\phi_i$ at random, $\phi_f$ will most probably be infinite,
with the function $\phi$ either increasing or decreasing logaritmically
as $\rho$ approaches $\rho_f$. Instanton solutions, if any, exist only 
for discrete $\phi_i$'s.
\vskip 1mm
Function $\phi$ for an instanton solution is 
confined to the barrier, and from its behaviour in the vicinity of
$\rho = 0$ and $\rho = \rho_f$ it is easily seen that it cannot
be located on one side of the barrier only. Indeed, it equals approximately
$\phi_i + V'_i \rho^2/8$ for $\rho \sim 0$ and $\phi_f + V'_f\
(\rho_f - \rho)^2/8$ for $\rho \sim \rho_f$, hence if it starts
and ends, say, on the near side of the barrier where $V' > 0$, it rises at
the beginning and falls at the end. Consequently, there must be at least
one turning point between $\rho = 0$ and $\rho = \rho_f$, and this turning 
point, or first of these turning points if more of them exist, must be located
on the far side of the barrier since $\dot \phi = 0$ and $\ddot \phi > 0$
there. As a result, the value $\phi_{top}$ at which $V$ reaches maximum
is crossed by the function $\phi$ at least once. If the total number
of crossings is $l$, one may call the solution 'CdL instanton of $l$th
order'.
\vskip 1mm
Equation for the function $\phi$ may be viewed as equation of motion of a
pointlike particle in the potential $- V$, with time-dependent friction 
coefficient $3\dot a/a$. The energy of the particle
\begin{equation}
E = \frac 12 \dot \phi^2 - V
\label{eq:En}
\end{equation}
obeys
\begin{equation}
\dot E = - \frac {3\dot a}a\ \dot \phi^2,
\label{eq:DEn}
\end{equation}
thus it decreases in friction regime, when $a$ increases
from zero to some maximum value, and increases in antifriction regime,
when $a$ returns to zero. For instanton solutions, the curve
$E (\phi)$ starts and ends on a wall of the barrier (which becomes a
well when we pass to the flipped potential), crossing the barrier $l$ times
if the instanton is of order $l$. For non-instanton solutions,
the curve looks the same except that it escapes to infinity
at the end. In the vicinity of an instanton solution there are 
non-instanton ones which either bounce back to infinity below the point
where the instanton solution stops, or slip forward to infinity above 
it, see fig.~\ref{fig:En}. The former solutions may be called undershootings
and the latter overshootings.
Continuity implies that in order to 'disentangle' the last turning 
point of a non-instanton solution, we must pass through a sequence containing 
necessarily an instanton solution having one turning point less. 
Furthermore, from a slight extension of the original Coleman argument valid
for an instanton in flat space \cite{col} it
follows that solutions starting sufficiently near to the true
vacuum are always 'short' overshootings escaping to infinity close to the 
starting point \cite{lys}. Putting this together we see that over
an instanton of $l$th order there must be a tower of instantons of orders
$l - 1,\ l - 2, \ \ldots$, possibly containing more instantons
of the same order. No matter how numerous the instantons are,
only that with the least action is of physical relevance. Presumably
this is the instanton of first order, at least for 'well-behaved'
potentials for which the complete tower of instantons contains one
representative of each order only and the starting point moves away
from $\phi_{top}$ as the order of instanton decreases.
\vskip 1mm
Introduce now a limit solution which is, at least half-and-half, shrinked 
to the top of the barrier, with $\phi = \phi_{top}$ inserted into the 
equation for $a$ and
the linearized expression for $V'$ inserted into the equation for $\phi$.
From the former equation we find $a = H^{-1} \sin (H\rho)$, where $H =
\sqrt{8\pi V_{top}/3}$, and inserting this into the latter equation we obtain
\begin{equation}
\Delta \ddot \phi + 3 H\ \mbox{cotan} (H\rho)\ \Delta \dot \phi = - \mu^2
\Delta\phi,
\label{eq:Eq1ap}
\end{equation}
where $\Delta \phi = \phi - \phi_{top}$ and $\mu = \sqrt{- V_{top}''}$. 
The 4D space is a 4-sphere with radius $H^{-1}$ and the 
equation for $\phi$ is the eigenvalue equation for Laplace operator 
on a 4-sphere. The equation assumes standard form if we pass from $\Delta 
\phi$ and $\rho$ to
$x = \Delta \phi/ \Delta \phi_i$ and $\varphi = H\rho$, and introduce
$\lambda = \mu^2/H^2$. Finite solutions exist only for $\lambda_l = l\
(l+3) = 0,\ 4,\ 10, \ldots$ and are of the form
\begin{equation}
x_l = 1,\ \cos \varphi, \ \frac 14 (5\cos^2 \varphi - 1), \ldots
\label{eq:lambda}
\end{equation}
Consider a 1-parameter class of potentials with $\lambda$ increasing from zero 
to infinity. For $\lambda = \lambda_l$ we have 'zero instanton 
of $l$th order' located at the top of the barrier (that is, dispersed 
infinitesimally around $\phi_{top}$); for other values of $\lambda$, we
have a non-instanton limit solution located at the top of the barrier on one 
side and escaping to $\phi = + \infty$ or $- \infty$
on the other side. Note that this soltion has a different asymptotics 
at $\rho \to \rho_f$ than the true solution, namely power-like instead of
logarithmic, thus it must be cut and matched to the true asymptotic solution 
at some $\rho$ to obtain a correct approximative solution. 
For $\lambda$ between $\lambda_l$ and $\lambda_{l + 1}$, 
the limit solution has the same number of turning points as $x_{l +
1}$. Indeed, when $x_{l + 1}$ deforms to $x$ with
a bit smaller $\lambda$, it does not turn to the opposite direction, see 
appendix \ref{ap:hfc}, and when one $x$ deforms to another, turning points
may not vanish or arise since $x$ may not have a maximum below zero or
a minimum above zero, see the argument against the localization of  
an instanton on one side of the barrier. This implies, by the 'disentanglement' 
argument, that for $\lambda$ between $\lambda_l$ and $\lambda_{l + 1}$
an instanton of $l$th order must exist somewhere at a finite distance 
from the top of the barrier, and consequently, that CdL instanton necessarily
exists for $\lambda > \lambda_1$. Inserting for $\lambda$ and $\lambda_1$
we obtain $\mu^2 > 4H^2$, or
\begin{equation}
-V'' > \frac{32\pi}3 V \ \ \mbox{at}\ \ \phi = \phi_{top}.
\label{eq:SC}
\end{equation}



\vskip 1mm
\section{Necessary condition}
\label{sec:NC}
\vskip 1mm
The condition from the previous section has been proposed in \cite{hmo} and 
discussed \cite{lys} in the context of new inflation, for potentials with 
a tiny barrier above a plateau with an almost zero slope. In fact,
the inequality in the cited articles differs from ours:
it reads $\mu^2 > 4H^2$, too, but with $H$ defined as $\sqrt{8\pi V_{f.v.}/3}$,
where $V_{f.v.}$ is the value of $V$ in false vacuum. However,
the considerations in both articles are approximative only, neclecting
the effect of small variations of the potential throughout the barrier
on the form of the 4D space, thus the difference between $V_{f.v.}$ and
$V_{top}$ plays no role in them.
In the first article it is claimed, with a brief comment referring to the
size of the bubble filled with true vacuum in flat space, that the 
validity of $\mu^2 > 4H^2$ is necessary and sufficient
for the existence of CdL instanton; in the second article
the necessary condition is supplemented by the assumption
that $V''$ decreases monotonously as one approaches $\phi_{top}$ from
the side of false vacuum (actually,
the authors require that $V''$ increases, but this is obviously an oversight)
and the proof of both necessary and sufficient condition is sketched.
In all considerations, the 4D space is supposed to have the metric of a 4-sphere.
As we have seen, the sufficient condition survives, with redefined $H$, also
in a theory without this simplification;
and a natural question arises whether this is possibly true about the
necessary condition as well. If it were, it would provide us with a simple 
algebraic criterion for the existence of CdL instantons, applicable
on a wide class of potentials including quartic. 
Unfortunately, we were not able to extend the argument of \cite{lys} to 
the exact theory. (In fact, the argument seems rather handwaving even in the 
approximation considered there.) We propose instead a condition which is
weaker in practical respect, but it still rules out a considerable portion 
of parameter space.
\vskip 1mm
In the next section it will turn out that the condition $\mu^2 > 4H^2$
is apparently necessary and sufficient at the same time for a rather
wide class of potentials, therefore it is perhaps useful to see why
this cannot be the case for any potential. An obvious counter-example is
a properly smoothed-out 'safe' potential: one takes a potential 
obeying the inequality $\mu^2 > 4H^2$ with large enough margin to allow 
for an instanton whose energy is high above $- V_{top}$, and smoothes out
the neighbourhood of $V_{top}$ so that the inequality $\mu^2 >
4H^2$ will not be valid anymore.
The constraint on $V''$ proposed in \cite{lys} is plausible at least in that
that it rules out such construction.
\vskip 1mm
To obtain the condition we are seeking for, let us pass from the equations 
for $\phi$ and $a$ to the equation for $E$, 
\begin{equation}
E'' = 8\pi (2E + 3V) + \frac {E'\left(\frac 23 E' + \frac 12 V'\right)}
{E + V}.
\label{eq:DDEn}
\end{equation}
Note that this equation is of second order, while the equation which one 
obtains from the system (\ref{eq:Eq12}) by elimination method is 
of third order: we have traded one order of differential equation for 
allowing the equation to contain the argument of unknown function.
The initial conditions are
\begin{equation}
E_i = V_i,\ \ E_i' = - \frac 34 V_i'
\label{eq:Eni}
\end{equation}
and the initial value of the second derivative of $E$ is
\begin{equation}
E_i'' = \frac {8\pi}3 V_i - \frac 12 V_i''.
\label{eq:DDEni}
\end{equation}
If we denote $\delta E = E + 3V/4$ and ${\cal V} = 8\pi V/3 + V''/4$, from
the expressions for $E_i'$ and $E_i''$ we find that
the function $\delta E'$ equals zero and its derivative equals
${\cal V}_i$ at $\phi = \phi_i$. For an instanton we have, in addition
to the two conditions at $\phi = \phi_i$, similar two conditions
at $\phi = \phi_f$; thus the function $\delta E'$ equals zero and
its derivative equals ${\cal V}_f$ there. Furthermore, if $\delta E'$
approaches zero outside the points $\phi_i$ and $\phi_f$,
its derivative tends to $8\pi (2E + 3V) + 3V''/4$, which is greater than
$8\pi V + 3V''/4 = 3{\cal V}$. Suppose ${\cal V} > 0$ throughout the
barrier. In such potential, $\delta E'$ for an instanton rises above zero 
at the beginning
and it should return to zero from below at the end, hence it should
cross zero from above somewhere in between; however, if it comes close to
zero at some point, its tangent becomes positive and it bounces
back. As a result, an instanton may exist only if there is a place somewhere
inside the barrier where the inequality ${\cal V} > 0$ is violated, that
is if
\begin{equation}
-V'' > \frac{32\pi}3 V\ \ \mbox{at some}\ \ \phi \ \ \mbox{inside the
barier}.
\label{eq:NC}
\end{equation}



\vskip 1mm
\section{Examples}
\label{sec:Ex}
\begin{figure}[h]
\centerline{\includegraphics[width=8cm]{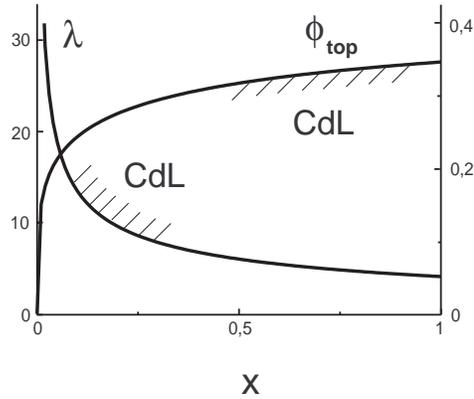}}
\caption{Allowed regions for CdL instanton in quartic potential}
\label{fig:Vqu1}
\end{figure}

\begin{figure}[h]
\centerline{\includegraphics[width=8cm]{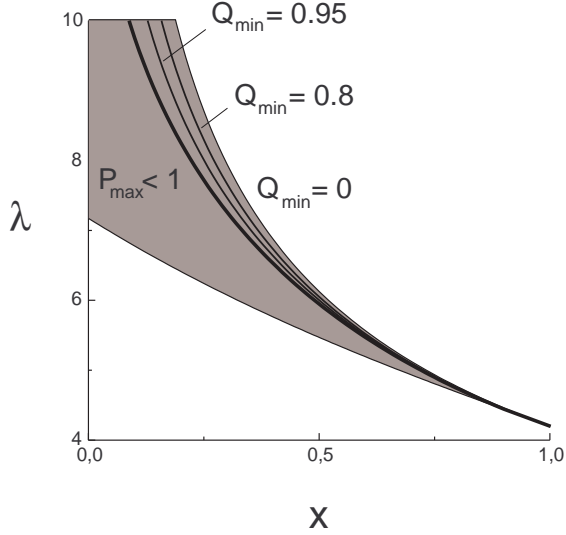}}
\caption{Topography of 'grey zone' for quartic potential}
\label{fig:Vqu2}
\end{figure}

\vskip 1mm
The first example we shall discuss is quartic potential
\begin{equation}
V = \frac 12 \phi^2 - \frac 13 \delta \phi^3 + \frac 14 \lambda \phi^4,
\label{eq:Vqu}
\end{equation}
with two parameters $\delta$ and $\lambda$ which are both supposed
to be positive. Usually, the first term is multiplied by $m^2$, 
so that $V$ contains one more parameter, the mass of the scalar field $m$. 
However, this parameter may be removed from the theory by the 
rescalings $a \to a/m$ and $\rho \to \rho/m$ in the equations for 
$\phi$ and $a$, supplemented by the rescalings $\delta \to m^2 \delta$ 
and $\lambda \to m^2 \lambda$ in the expression for $V$.
\vskip 1mm
The quartic potential is of the type we are interested in in a rather narrow
range of parameters, for $\delta$ ranging from 
$\delta_m = 2 \sqrt{\lambda}$ to $\delta_M = 3 \sqrt{\lambda/2} \doteq 1.06 
\delta_m$. This is discussed, along with other properties of $V$,
in appendix \ref{ap:Vqu}. In what follows, we use instead of $\delta$
its linear transform
\begin{equation}
x = \frac {\delta - \delta_m}{\delta_M - \delta_m},
\label{eq:parx}
\end{equation}
rearranging the parameter space into a half-infinite strip of unit width.
\vskip 1mm
Let us investigate the possibility of the existence of CdL instanton
in quartic potential.
Denote $\lambda_N$ and $\lambda_S$ the limit values of $\lambda$
at which the necessary and sufficient condition for the
existence of CdL instanton are valid. The two $\lambda$'s correspond
to two particular configurations of the curves $\hat V = 32\pi V/3$
and $- V''$: at $\lambda_S$ the latter curve crosses the former at
its peak, and at $\lambda_N$ the latter curve becomes tangent to the
former at some point, lying
below it everywhere outside that point. Denote, furthermore,
$\phi_S$ the value of $\phi_{top}$ for $\lambda = \lambda_S$. 
In appendix \ref{ap:Vqu}, the dependence of $\lambda_S$ and $\phi_{S}$
on $x$ is given analytically, see equations (\ref{eq:lSC}),
(\ref{eq:phSC}) and
(\ref{eq:parX}), and the dependence of $\lambda_N$ on $x$ is expressed in
parametric form, see equations (\ref{eq:osc}) and the discussion below. The
first two functions, together with allowed regions of $\lambda$ and
$\phi_{top}$, are depicted in fig.~\ref{fig:Vqu1}.
The allowed regions are shaded at the boundary and denoted by the
corresponding variable, and two sets of tic labels are given at
the vertical axes, those referring to $\lambda$ on the left
and those referring to $\phi_{top}$ on the right.
As we can see, the values of $\lambda$ are bounded from below while
the values of $\phi_{top}$ are bounded from above.
The domain $\lambda_N < \lambda < \lambda_S$, with contour lines
characterizing the behaviour of the solutions of instanton equations,
is depicted in fig.~\ref{fig:Vqu2}.
This is the 'grey zone' of the parameter space in which we cannot decide
whether the potential admits an instanton or not but numerically. Sample
calculations reveal no instanton solutions there, but this of course
does not rule out the possibility that they exist.
The character of the solutions may be explored systematically by
defining an appropriate global characteristic of the solutions and 
drawing its contour lines. For a potential from the 'grey
zone', the curve $-V''$ reaches out of the curve $\hat V$ only in some
interval $(\phi_A, \phi_B)$ between $\phi_{top}$ and the true vacuum at
$\phi = 0$, shrinking from maximum size to a point as $\lambda$ decreases 
from $\lambda_S$ to
$\lambda_N$. Consider solutions to the instanton equations starting on 
the true-vacuum side of $\phi_{top}$ for which the function $\delta E'$ 
is negative somewhere; these are all solutions starting in the interval
$(\phi_A, \phi_B)$, since for them the function $\delta E'$ becomes negative
immediately after the start, and possibly some solutions starting 
before $\phi_A$, since for them this function may become negative
later. If the potential does not admit instantons,
the function $\delta E'$ crosses zero again at some $\phi_{lim}$ and escapes
to infinity after. Since the solutions do not come to a halt at
$\phi_{lim}$, the quantity $\Delta E = E + V$ is positive there, 
$\Delta E_{lim} > 0$. If, on the other hand, the potential does admit
instantons, they belong to the class of solutions in question; and among
them there is an instanton of first order for which the function $\delta E'$
stops after returning to zero on the far side of $\phi_{top}$,
so that $\phi_{lim} > \phi_{top}$ and $\Delta E_{lim} = 0$. Define
\begin{equation}
P = \frac {\phi_{lim}}{\phi_{top}}, \ \ Q = \left \{ \begin{array} {l}
\Delta E_{lim}/(E_{lim} + V_{top}) \ \mbox{if} \ P \ge 1 \\
1 \ \mbox{if} \ P < 1 \\
\end{array} . \right.
\label{eq:PQ}
\end{equation}
As follows from previous discussion, for potentials admitting no instantons 
$P_{max}$ may be $> 1$ as well as $\le 1$ and $Q_{min}$ is necessarily
positive, while for potentials that do admit instantons $P_{max} > 1$ 
and $Q_{min} = 0$. By continuity argument, if we are close enough to a domain
in parameter space in which instantons are allowed, $P_{max}$ must exceed 1,
and as we approach the bordeline of this domain from outside, $Q_{min}$ must
decrease from some moment on until it reaches zero at the borderline. 
In fig.~\ref{fig:Vqu2},
the thick line corresponds to $P_{max} = Q_{min} = 1$ and the thin
lines above it, in the strip where $P_{max} > 1$, correspond
to values of $Q_{min}$ given next to them. The form of the lines
strongly suggests that as we cross the strip, $Q_{min}$ decreases
from 1 to 0 monotonously, hence no instantons occur in the 'grey zone'.
\vskip 1mm
Quartic potential may be used to demonstrate how difficult it is
to obtain open inflation in a 'natural' way. A single-field open inflation 
requires an instanton solution whose limit value of $\phi$ on the
true-vacuum side of $\phi_{top}$ is greater than about 3; and 
from a simple qualitative analysis it follows \cite{{lme},{toy}} that 
for quartic potential no such solution exists. Our computation confirms 
this. Indeed, the allowed region of $\phi_{top}$ in fig.~\ref{fig:Vqu1} is
located deeply under the line $\phi_{top} = 3$, and as long as
one regards the whole 'grey zone' as forbidden for the instantons,
this region contains
all admissible values of $\phi_{top}$. Consequently, the values
of $\phi_{top}$ for potentials that do admit instantons are significantly
less than 3, and the same is true of any value of $\phi$ that may be
reached by an instanton solution on the true-vacuum side of $\phi_{top}$.
\vskip 1mm
Let us proceed to our second example, a superposition of quadratic and
Breit - Wigner type potential 
\begin{equation}
V = \frac 12 \phi^2 \left[ 1 + \frac {\alpha^2}{\beta^2 + (\phi - v)^2}
\right].
\label{eq:VLin}
\end{equation}
Again, we have scaled away the mass $m$. This potential has been introduced
by Linde in his toy model of open inflation \cite{toy}, with the mass
set to the value $m \simeq 10^{-6}$ in order to get the proper size
of density fluctuations. For $\beta$ sufficiently small,
$V$ has a peak centered near $\phi = v$ with the width $\beta$ and relative
overheight $k = \alpha^2/\beta^2$. To illustrate the properties of such 
potential, the values $v = 3.5$, $k = 0.5$ and $\beta = 0.1$ have been 
chosen in \cite{toy}.
To observe the constraint on the potential coming from the scenario of open 
inflation (provided the peak is narrow, which is necessarily the case if an
instanton is to occur at all), one has to take $v \gtrsim 3$;
and a value at the lower end of this range has to be chosen
if one wishes to obtain the cosmological parameter $\Omega$ not too
close to 1. Denote $k_N$ and $k_S$ the limit values of $k$ for which the
necessary and sufficient condition of the existence of CdL instanton
is fulfilled. In appendix \ref{ap:VL} it is shown 
that if $v$ is large enough to allow for open inflation, these quantities
approximately coincide, and they increase from zero to infinity 
as $\beta$ increases from zero to $B = \sqrt{3/(16\pi)} \doteq 0.24$, 
switching from linear to quadratic dependence on $\beta$ near $b = B^2/v$, 
$b \doteq 0.017$ for $v = 3.5$. The corresponding formulas are summarized
in equation (\ref{eq:kap}). Between $k_S$ and $k_N$
there should be a 'grey zone'. In the first
approximation this zone is absent; however, its size may be
deduced from the corrections to $k_S$ and $k_N$, see equations (\ref{eq:DB})
and (\ref{eq:grey1}) for $\beta \gg b$ and equation
(\ref{eq:grey2}) for $\beta \ll b$. The zone turns out to be quite narrow;
for example, for values of $v$
and $\beta$ used in \cite{toy} one obtains $k_S = 0.23$ and $k_N = 0.20$.
The possible existence of CdL instanton in the 'grey zone' could now be
investigated by drawing similar contour lines as for quartic
potential. However, for $\beta \gg b$ one may determine the character of
the 'grey zone' on the basis of the results for quartic potential only. The
point is that in this part of 'grey zone' the curve $- V''$ exceeds
the curve $\hat V$ only in a tiny interval of $\phi$ close to
$\phi_{top}$, like for quartic potential with $x$ close to 1, and both
curves may be approximated by inverted parabolas throughout this interval,
again like for quartic potential with $x$ close to 1. Consequently, since
the interval under consideration plus a small adjacent region is all what is
relevant for the problem of the existence of CdL instanton, and since all
the 'grey zone' for quartic potential is presumably forbidden, the 'grey zone'
for Linde potential with $\beta \gg b$ should be forbidden, too. 
On the other hand, for $\beta \ll b$ the zone is so narrow that it is
of no relevance at all, in view of the uncertainties in the form of the 
potential in any not-exactly-toy model.



\vskip 1mm
\section{Conclusion}
\vskip 1mm
We have shown that if the inequality $- V'' > 32\pi V/3$ holds at the
top of the barrier, the potential admits CdL instanton, and if this
inequality is valid at least somewhere throughout the barrier, the
potential {\it may} admit CdL instanton. Otherwise only HM instanton exists.
(HM instanton is by definition trivial solution to equations
(\ref{eq:Eq12}), which in our notations reads $\phi = \phi_{top}$
and $a = H^{-1} \sin (H\rho)$.) The necessary condition for the existence
of CdL instanton does not restrict the form of the potential in advance,
like the constraint on $V''$ in \cite{lys}. However, this is achieved 
at the price that the condition is weaker than the sufficient condition
for any but very specific potentials,
not allowing for the algebraic solution of the problem even in the
simplest case of quartic potential. To obtain a sharp borderline
between the regions in parameter space in which vacuum decays in the way
described by CdL and HM instanton, one may use the method of contour lines
outlined in section \ref{sec:Ex}. As we have seen, by applying this method
on quartic potential one finds that the sufficient condition is most
probably also necessary, thus the existence of CdL instanton depends
uniquely on the form of the potential at the top of the barrier.
The same is true, at least for potentials whose barrier is not too narrow,
about Linde potential. An open question remains what is the largest class
of potentials with this property.


\vskip 5mm
\noindent
{\it Acknowledgements.} This work was supported by the grant VEGA 1/0250/03.



\vskip 1mm
\appendix 
\section{Unbounded solutions to the eigenvalue problem}
\label{ap:hfc}
\setcounter{equation}{0}
\renewcommand{\theequation}{A-\arabic{equation}}
\vskip 1mm
The eigenvalue equation
\begin{equation}
\ddot x + 3\ \mbox{cotan} \varphi\ \dot x + \lambda x = 0,
\label{eq:Eqx1}
\end{equation}
where the dot denotes differentiation with respect to $\varphi$, transforms
under the substitution $z = \sin^2 (\varphi/2)$ into an equation for
(nondegenerate) hypergeometric function
\begin{equation}
z(1 - z)\ x'' + 2(1 - 2z)\ x' + \lambda x = 0,
\label{eq:Eqx2}
\end{equation}
where the prime denotes differentiation with respect to $z$. This yields
\begin{equation}
x = F \left( \frac 32 + \Lambda, \ \frac 32 - \Lambda, \ 2;
\ \sin^2 \frac \varphi 2 \right),\ \
\Lambda \equiv \sqrt{\frac 94 + \lambda},
\label{eq:x}
\end{equation}
and with the help of the identity 
$$F(a,\ b,\ c;\ z) = (1 - z)^{c - a - b} F(c - a,\ c - b,\ c;\ z)$$
we find that $x$ diverges quadratically as $\varphi$ approaches $\pi$,
$x \simeq k (\pi - \varphi)^{-2}$, with the constant of proportionality
\begin{equation}
k = 4F \left( \frac 12 + \Lambda, \ \frac 12
- \Lambda,\ 2;\ 1 \right) = - \frac {4 \cos (\Lambda \pi)}
{(2 + \lambda)\pi}.
\label{eq:xap}
\end{equation}
The function $k (\lambda)$ oscillates around zero with nods at $\lambda =
\lambda_l$ and is negative between $\lambda_0$ and $\lambda_1$.
Consequently, if $\lambda$ is from the interval
($\lambda_l,\ \lambda_{l + 1})$, $x$ diverges to $- \infty$ and $+ \infty$
for even and odd $l$ respectively, and since $x_{l+1}$ decreases when
approaching its endpoint for even $l$ and increases for odd $l$, no
turning point arises when $x$ disconnects from $x_{l + 1}$.



\vskip 1mm
\appendix 
\renewcommand{\thesection}{B}
\section{Properties of quartic potential}
\label{ap:Vqu}
\setcounter{equation}{0}
\renewcommand{\theequation}{B-\arabic{equation}}
\vskip 1mm
Quartic potential has a minimum equal to zero at $\phi = 0$ and
for $\delta > \delta_m = 2\sqrt{\lambda}$ it has another two extremes at
\begin{equation}
\phi_{\pm} = \frac \delta {2\lambda} \pm \sqrt{\frac {\delta^2} {4\lambda^2}
- \frac 1\lambda};
\label{eq:ppm}
\end{equation}
furthermore, for $\delta < \delta_M = 3\sqrt{\lambda/2}$ it is nonnegative 
everywhere, with only zero at $\phi = 0$. Consequently, for $\delta_m <
\delta < \delta_M$ the potential has the desired form, with
true vacuum at $\phi = 0$, false vacuum at $\phi = \phi_+$ and the top
of the barrier at $\phi = \phi_-$. Let us compute $\lambda_S$
and $\phi_S$ for such potential. In present notations, $\lambda_S$
is given by the equation $\mu_-^2(\lambda_S) = 4H_-^2(\lambda_S)$, where
$H_\pm^2 = 8\pi V_\pm/3$ and $\mu_\pm^2 = |V_\pm''|$, and $\phi_S$
equals $\phi_-(\lambda_S)$. It holds
$$H_\pm^2 = \frac {2\pi}{9\lambda} \frac {(1 \pm X)(1 \mp 3X)}{(1 \mp X)^2},
\ \ \mu_\pm^2 = \frac {2X}{1 \mp X},\ \ X \equiv \sqrt{1 - \frac{4\lambda}
{\delta^2}},$$
hence the equation for $\lambda_S$ yields
\begin{equation}
\lambda_S = \frac {4\pi}9 \frac {(1 - X)(1 + 3X)}{X(1 + X)},
\label{eq:lSC}
\end{equation}
and if we express $\phi_-$ in terms of $\lambda$ and $X$ and exploit the
above equation, we find
\begin{equation}
\phi_S = \frac 32 \sqrt {\frac X{\pi (1 + 3X)}}.
\label{eq:phSC}
\end{equation}
The parameter $X$ depends on the parameter $x$, running from 0 to 1/3
as $x$ runs from 0 to 1,
\begin{equation}
X =  \sqrt{1 - \frac 1{(1 + \kappa x)^2}},\ \ \kappa \equiv \frac {\delta_M}
{\delta_m} -1.
\label{eq:parX}
\end{equation}
Let us proceed to $\lambda_N$. After the rescalings $\phi \to \phi/
\sqrt{\lambda}$ and $\delta \to \sqrt{\lambda} \delta$ we have
$$\hat V = \frac {32\pi}{3\lambda} \phi^2 \left(\frac 12 - \frac 13 \delta
\phi + \frac 14 \phi^2 \right),\ \ - V'' = - 1 + 2 \delta \phi - 3 \phi^2,$$
so that if we change $\lambda$ leaving $\delta$ unchanged, the curve $- V''$
remains the same while the curve $\hat V$ scales by a factor inversely
proportional to $\lambda$.
Note that $\delta$ unchanged now means $x$ unchanged, too, since we have
replaced $\delta$ by $\delta_{new} = \delta/\sqrt{\lambda} = 2(1 + \kappa
x)$. The curves $- V''$ and $\hat V$ are both concave in the relevant
interval in which $- V''$ exceeds zero; consequently, if $\lambda$
decreases from infinity to zero, so that $\hat V$ transforms from a
segment on horizontal axis to an infinitely high peak, the part of
$- V''$ reaching out of $\hat V$ shrinks from the whole curve $- V''$
in the interval of interest to a point, and then the two curves disconnect.
As a result, for each $\delta$ we have unique $\lambda$ for which the
curves $- V''$ and $\hat V$ are tangent, and since the former curve is 
located below the latter everywhere outside the oscular point, this $\lambda$ 
equals $\lambda_N$. Denote $\phi_{osc}$ the value of $\phi$ at the oscular 
point. For given $\delta$, $\phi_{osc}$ is determined, simultaneously with
the critical $\lambda$ for which the oscular point exists, by the equations
\begin{equation}
\hat V_{osc} = - V_{osc}'',\ \ \hat V_{osc}' = - V_{osc}''', 
\label{eq:osc}
\end{equation}
and since the critical $\lambda$ is just $\lambda_N$, these equations
may be viewed as a parametric definition of the function $\lambda_N(\delta)$.
If we return from rescaled $\delta$ and $\phi$ to the original ones, the 
equations become linear in $\delta$ and $\lambda_N$ and solve trivially
for any given $\phi_{osc}$. The only remaining task is to determine
the limits of $\phi_{osc}$. The upper limit is reached at $x = 1$ and
coincides with the value of $\phi_S$ there, which is $\sqrt{3/(8\pi)}
\doteq 0.35$; the lower limit is reached at $x = 0$ and if we solve equations
(\ref{eq:osc}) numerically we find that it equals about 0.22.



\vskip 1mm
\appendix 
\renewcommand{\thesection}{C}
\section{Properties of Linde potential}
\label{ap:VL}
\setcounter{equation}{0}
\renewcommand{\theequation}{C-\arabic{equation}}
\vskip 1mm
The extremes of Linde potential outside $\phi = 0$ are given by
\begin{equation}
\left( 1+ x^2 \right)^2 = k \left( \frac x\sigma - 1\right),
\label{eq:exL}
\end{equation}
where $x = (\phi - v)/\beta$ and $\sigma = \beta/v$. Suppose $\sigma \ll
1$. (This assumption will be justified shortly). Two asymptotic regimes
may be distinguished: if $k \gg \sigma$, the extremes are widely 
separated in the variable $x$, and if $k \doteq \kappa \sigma$, where 
$\kappa$ is a constant to be determined, the extremes are close to each
other. In the former limit, the maximum is located close to zero at the 
point $x = (1 + 1/k)\sigma$ and the minimum is located far from 1 at
the point $x = (k/\sigma)^{1/3}$; in the latter limit, both extremes 
are located close to the point $x = a$ at which they merge if $k \to 0$, 
$\sigma \to 0$ and $k/\sigma = \kappa$. 
By solving the equations $V' = V'' = 0$ with $k \to 0$, $\sigma \to 0$ and 
finite $k/\sigma$ we obtain $a = 1/\sqrt{3}$ and $\kappa = 16/(3\sqrt{3})$.
Let us find approximate expressions for $k_{S,N}$ as functions of $\beta$
and $v$. If $k_{S,N}$ are significantly greater than
$\sigma$, they are both approximately given by the equation $- V_0'' 
= \hat V_0$ where the index 0 refers to the point $x = 0$. (By definition, 
$k_S$ and $k_N$ are given by the equation $- V'' = \hat V$ at the points 
$x_{top}$ 
where $\hat V$ has maximum and $x_{osc}$ where $\hat V$ is tangent to
$- V''$, however, both points are close to zero in the limit considered.)
It holds $\hat V_0 = 16\pi v^2(1 + k)/3$ and $- V_0'' \doteq k/\sigma^2 = k
v^2/\beta^2$, hence $k_{S,N}$ are both approximately equal to $\beta^2/(B^2 
- \beta^2)$, where $B = \sqrt{3/(16\pi)}$. The requirement that this
expression is much greater than $\sigma$ yields $\beta \gg b = B^2/v$.
Denote $k_{lim}$ the value of $k$ at which the two extremes of the potential 
outside $\phi = 0$ merge, and under which the theory becomes inapplicable since 
no potential barrier exists. In the interval $\beta \ll b$ both $k_{S,N}$
must be close to $k_{lim}$ which equals approximately
$k_{lim}^{(0)} = \kappa \sigma = \kappa \beta/v$ there; $k_S$ must be 
close to $k_{lim}$ since then the curve
$- V''$ does not reach too high at the maximum of $\hat V$, being equal to zero 
at the inflex point of $\hat V$ if $k = k_{lim}$, and $k_N$ must be close to
$k_{lim}$ (actually, it coicides with $k_{lim}$) since it may not exceed  
$k_S$. Putting this together we obtain
\begin{equation}
k_{S,N} \doteq \left \{ \begin{array} {l}
\beta^2/(B^2 - \beta^2)\ \mbox{if}\ \beta \gg b \\
\kappa \beta/v\ \mbox{if}\ \beta \ll b\\
\end{array} . \right.
\label{eq:kap}
\end{equation}
The upper limit of $\beta$ is $B$, so that the upper limit of $\sigma$ is $B/v$; 
and since this number is less than 0.08 for $v > 3$, our approximation works in 
the whole relevant range of parameters, if only as a qualitative estimate
at its edge. Let us now discuss the corrections to $k_{S,N}$. In the interval
$\beta \gg b$, we start from writing down the expansions of $\hat V$ and $- V''$ 
up to the second order in $x$,
$$\hat V = \frac {16\pi}3 v^2 (1 + k) \left(1 + 2\sigma x -\frac k{1+k} x^2
\right),\ \ - V'' = \frac k{\sigma^2} \left(1 + 6\sigma x - 6 x^2
\right) - 1.$$
If we evaluate $k_S$ from the condition that the two parabolas intersect at
the top of the first of them, and $k_N$ from the condition that the two
parabolas are tangent, we obtain the previous expression with 
$B$ replaced by $B_{S,N} = B$ plus corrections of order $\sigma^2$. 
The shift of $B_S$ with respect to $B_N$ is
\begin{equation}
\Delta B = - \frac {9(1 + k)(2 + k)^2}{2k^2(6 + 5k)}\ \sigma^2 B.
\label{eq:DB}
\end{equation}
The curves $k_S$ and $k_N$ differ the most at $\beta$ not too
close to $B$, where both $k_S$ and $k_N$ are small and the correction to $B_S$ 
dominates over the correction to $B_N$. Consequently, a good approximation
of $k_S$ and $k_N$ is obtained for any admissible $\beta$ if one puts 
$B_N = B$, $B_S = B + \Delta B(k_S)$; in this way the values of $k_S$ and
$k_N$ cited in the text have been calculated. Furthermore, from the asymptotics 
of $\Delta B$ for $k \gg 1$ and $k \ll 1$ we find that, as we pass from
$k_N$ to $k_S$, the upper limit of $\beta$ and the values of $k$ at
$\beta \ll B$ become shifted by
\begin{equation}
\Delta \beta_{lim} = - \frac 9{10} \frac {B^3}{v^2},\ \ \Delta k =
\frac {6B^2}{v^2}.
\label{eq:grey1}
\end{equation}
In the interval $\beta \ll b$, we may evaluate $k_S$ from the equations 
$V_{top}' = 0$ and $- V_{top}''= \hat V_{top}$, assuming $x_{top}$ is 
close to $a$ and $k_S$ is close to $k_{lim}^{(0)}$. In this way
we find that $k_S$ differs from $k_{lim}$ by a term proportional to
$\beta^3$ only, while $k_{lim}$ itself differs from $k_{lim}^{(0)}$ by a
term proportional to $\beta^2$. A similar procedure for $k_N$ yields
nothing, which is understandable: as we shift the extremes to each other,
the curve $\hat V$ relaxes to the parabola $16\pi \phi^2/3$, while retaining
a ripple near $\phi = v$ whose slope varies rapidly; and because of this
ripple, the curve $- V''$ acquires a sharp peak sticking out of $\hat V$.
Consequently, $k_N$ in the interval $\beta \ll b$ coincides with $k_{lim}$
and $\Delta k$ with $k_S - k_{lim}$. Explicit calculation yields
\begin{equation}
\Delta k = \frac 19 \frac{\kappa \beta^3}{v b^2}.
\label{eq:grey2}
\end{equation}



\end{document}